\documentclass[floatfix,aps,prd,twocolumn,showpacs,notitlepage,eqsecnum,
superscriptaddress,nofootinbib]{revtex4-1}

\usepackage[centertags]{amsmath} \usepackage{amssymb} \usepackage{latexsym}
\usepackage{enumerate} \usepackage{graphicx} \usepackage{mathrsfs}
\usepackage[colorlinks]{hyperref} \usepackage{stmaryrd} \usepackage{import}
\usepackage{tensor} \usepackage[usenames,dvipsnames]{xcolor} \usepackage{bm}
\usepackage{multirow} \usepackage{breakurl} \usepackage{float}
\usepackage{verbatim} \usepackage{mathrsfs}
\usepackage{orcidlink}

\allowdisplaybreaks[1]

\definecolor{CiteColor}{rgb}{0,0.5,0} \hypersetup{citecolor=CiteColor}
\definecolor{RefColor}{rgb}{0.55,0,0} \hypersetup{linkcolor=RefColor}
\definecolor{darkgreen}{rgb}{0.2,0.7,0.2}

\providecommand{\e}[1]{\ensuremath{\times 10^{-#1}}}
\newcommand{\msp}{\phantom{-}}

\begin{document}

\title{New self-force method via elliptic partial differential equations for Kerr inspiral models}

\author{Thomas Osburn$\,$\orcidlink{0000-0003-2747-3994},}
\email{tosburn@geneseo.edu}
\affiliation{Department of Physics and Astronomy, 
State University of New York at Geneseo, New York 14454, USA}
\affiliation{Department of Physics and Astronomy, 
University of North Carolina, Chapel Hill, North Carolina 27599, USA}
\author{Nami Nishimura}
\email{nnishimu@umd.edu}
\affiliation{Department of Physics and Astronomy, 
State University of New York at Geneseo, New York 14454, USA}
\affiliation{Department of Physics, 
University of Maryland, College Park, Maryland 20742, USA}

\begin{abstract}
We present a new method designed to avoid numerical challenges that have impeded calculation of the Lorenz gauge self-force acting on a compact object inspiraling into a Kerr black hole. This type of calculation is valuable in creating waveform templates for extreme mass-ratio inspirals, which are an important source of gravitational waves for the upcoming Laser Interferometer Space Antenna mission. Prior hyperbolic partial differential equation (PDE) formulations encountered numerical instabilities involving unchecked growth in time; our new method is based on elliptic PDEs, which do not exhibit instabilities of that kind. For proof of concept, we calculate the self-force acting on a scalar charge in a circular orbit around a Kerr black hole. We anticipate this method will subsequently facilitate calculation of first-order Lorenz gauge Kerr metric perturbations and self-force, which could serve as a foundation for second-order Kerr self-force investigations.
\end{abstract}

\maketitle

\section{Introduction}

Gravitational wave observations by the Laser Interferometer Gravitational-Wave Observatory (LIGO) and Virgo have significantly enhanced astrophysics research~\cite{Abbott_2019,Abbott_2021,Abbott_2022,LIGO_gr,LIGO_pop}. Expansion of the current-generation network is underway through inclusion of the Kamioka Gravitational Wave Detector~\cite{KAGRA1,KAGRA2} and likely future inclusion of LIGO-India~\cite{Saleem_2021}. Recently, the gravitational wave community has directed further attention toward the next generation of detectors. Among the most valuable of these future detectors is the Laser Interferometer Space Antenna (LISA)~\cite{LISA}, which will probe millihertz gravitational waves for the first time. Exploring this region of the gravitational wave spectrum is expected to unveil new sources such as supermassive black hole binaries and extreme mass-ratio inspirals (EMRIs). EMRIs involve a stellar-mass compact object (mass~$\mu~\sim~10\,M_{\odot}$) slowly inspiraling into a massive black hole (mass~$M~\sim~10^6 M_{\odot}$). Here we focus on theoretical EMRI calculations with applications to gravitational wave astronomy. If quantitative EMRI models are able to achieve sufficient accuracy and realism, LISA observations of EMRIs will successfully probe strong-field gravity with unprecedented precision~\cite{EMRI}. This work investigates and overcomes technical obstacles that have inhibited prior inclusion of certain valuable EMRI features in theoretical models.

General relativity is the prevailing theory to describe compact binary systems like EMRIs. Successful identification of a valid and/or suitable approach follows from analysis of appropriate properties of the two-body system. The three main approaches are post-Newtonian theory~\cite{Blanchet_2014}, numerical relativity~\cite{Brugmann_2018}, and black hole perturbation theory~\cite{Pound_2021}. Post-Newtonian theory is favorable for large separations. Numerical relativity is favorable for comparable masses with small separations. Black hole perturbation theory is favorable for extreme mass ratios with small separations. We use ``favorable" because there are regions of overlap in parameter space between different approaches, but for some systems there may only be one viable approach. EMRIs are an example of a such a system; the mass ratios are sufficiently extreme to defeat numerical relativity (mass ratio $\epsilon \equiv \mu/M \sim 10^{-5}$), and the late phase of the inspiral is too relativistic for post-Newtonian theory. Therefore, black hole perturbation theory seemingly is the only favorable option to accurately describe EMRI dynamics (other EMRI frameworks, such as effective one-body theory~\cite{Buonanno_1999,Antonelli_2020}, adopt appropriate results from black hole perturbation theory). With the black hole perturbation theory approach, the larger binary component provides the leading term in an expansion of the spacetime metric in powers of the small mass ratio ($\epsilon$); higher-order terms describe dynamical two-body interactions. 

In applying black hole perturbation theory to develop comprehensive EMRI waveform templates, perhaps the single most important quantity to calculate accurately is the waveform phase. During the sustained inspiral, approximately $\epsilon^{-1} \sim 10^5$ rad of phase accumulate through merger. To achieve accurate parameter estimation and precisely test general relativity, the phase evolution will need to be calculated to within an absolute error significantly smaller than $\sim 1$ rad. Phase calculations follow from determining the small body's orbital motion throughout the inspiral. The self-force~\cite{Mino_1997,Quinn_1997,Poisson_2004}, a mechanism where the small body's gravitational perturbation backreacts onto itself, drives orbital decay. Therefore, the accuracy of self-force calculations used during orbital evolution will determine the crucial phase accuracy. To achieve LISA-motivated accuracy goals, the conservative part of the self-force needs to be calculated through first order in~$\epsilon$~\cite{Pound_2005}, while the dissipative part of the self-force needs to be calculated through second order in~$\epsilon$~\cite{Hinderer_2008}.

The need for second-order calculations presents considerable challenges. One manifestation of general relativity's nonlinearity is how the second-order metric perturbation is sourced by products involving the first-order metric perturbation and itself (and its derivatives). Considering that the first-order metric perturbation becomes infinite approaching the small body, great care must be applied in constructing and implementing a well-behaved second-order source. Recently, a decade of dedicated effort has culminated in the earliest (and, as of now, only) second-order EMRI model~\cite{Pound_2020,Warburton_2021,Wardell_2021}; facing seemingly atrocious complications associated with the second-order self-force problem, that second-order effort considered the Schwarzschild quasicircular case to maximize simplicity (presumably). One feature that may have aided successful second-order calculations was usage of the Lorenz gauge, where challenges related to the second-order problem are better understood~\cite{Pound_2012,Pound_2014,Miller_2021}. These second-order calculations require, as input, an accurate first-order metric perturbation (and derivatives) in a sufficiently regular gauge such as the Lorenz gauge. For Schwarzschild metric perturbations, the first-order Lorenz gauge problem has been conquered by performing a tensor spherical harmonic decomposition~\cite{Barack_2002,Detweiler_2004,Barack_2005,Barack_2007,Berndtson_2007,Barack_2009,Akcay_2011,Warburton_2012,Akcay_2013,Osburn_2014,Hopper_2015,Wardell_2015,Osburn_2016,Warburton_2017,van_de_Meent_2018,Akcay_2020,McCart_2021}. Achieving accessibility of Lorenz gauge metric perturbations in more realistic and/or comprehensive scenarios would be a considerable advancement. One example of a more realistic/comprehensive scenario is when the bodies (one or both) are spinning. Spin for the larger body is achieved by calculating perturbations of Kerr spacetime. Unfortunately, past efforts to calculate the first-order Lorenz gauge Kerr self-force have faced technical challenges and were not entirely successful. 

The Lorenz gauge Kerr perturbation equations are not known to be directly separable into multipole modes. Accordingly, past work has approached the problem of Kerr perturbations in Lorenz gauge by separating the $\phi$ dependence into $m$-modes and solving hyperbolic partial differential equations (PDEs) involving $t$, $r$, and $\theta$ derivatives. Although this strategy is enticing, it was not entirely successful because instabilities affecting the $m=0$ and $m=1$ modes caused the numerical solution to grow without bound proportionally to time~\cite{Dolan_2013}. Lorenz gauge approaches seems to transcend specific numerical methods, appearing with $u$-$v$ coordinates~\cite{Barack_2007}, $t$-$r_*$ coordinates~\cite{Dolan_2013}, and gauge driver methods~\cite{Thornburg_2018}. Recently, a technique based on orthogonalization in solution space has been successful in taming these numerical instabilities for the Schwarzschild case, and it is plausible that this technique will also be successful for Kerr~\cite{Thornburg_2020,Thornburg_2022}. In this work, we introduce and implement a new method that avoids problematic instabilities by entering the frequency domain and solving elliptic PDEs with $r$ and $\theta$ derivatives.

Rather than grappling with nonseparable PDEs (as we are proposing), many popular approaches to the Kerr perturbation problem have begun with the separable Teukolsky equation~\cite{Teukolsky_1972,Teukolsky_1973}. The Teukolsky approach has been particularly fruitful for radiation gauge metric perturbations~\cite{Chrzanowski_1975,Wald_1978,Kegeles_1979,Lousto_2002,Ori_2003,Keidl_2007,Keidl_2010,Shah_2011,Pound_2014b,van_de_Meent_2015,Merlin_2016,van_de_Meent_2017,Barack_2017}, for which the Kerr first-order self-force has been calculated in a variety of configurations~\cite{Shah_2012,Isoyama_2014,Colleoni_2015,van_de_Meent_2016,van_de_Meent_2017b,van_de_Meent_2018b,Thornburg_2020b,Lynch_2022}. However, radiation gauges are not especially well behaved, which may be why existing second-order EMRI models have not used radiation gauges; although, ongoing developments toward radiation gauge based second-order calculations are being made~\cite{Green_2020,Toomani_2021}. Recently, a promising Teukolsky based approach for vacuum Lorenz gauge metric reconstruction was discovered~\cite{Dolan_2022}; such a method may become the best of both worlds upon future generalization to nonvacuum scenarios. Although this work pursues a non-Teukolsky based approach, we believe these alternate pathways toward second-order Kerr perturbations will lead to valuable cross-verification of results in the future.

The new method we present and implement here is designed to calculate Kerr perturbations directly in Lorenz gauge while avoiding problematic instabilities encountered in prior work. Our method leverages the two Kerr Killing vectors by separating the $\phi$ and $t$ variables. Entering the frequency domain is a key element of our strategy; the associated elliptic PDEs do not involve instabilities like those encountered in the prior hyperbolic PDE formulation~\cite{Dolan_2013}. As proof of concept, we first apply our method to calculate the self-force on a scalar charge in a circular equatorial orbit around a Kerr black hole. The case involving Lorenz gauge Kerr metric perturbations will be pursued in subsequent work. It is somewhat strange to approach the Kerr scalar self-force in the frequency domain via PDEs because, unlike the gravitational case, the scalar field equation is directly separable into spheroidal harmonics~\cite{Warburton_2010,Warburton_2011,Warburton_2015,Nasipak_2019,Nasipak_2021} as the $s=0$ instance of the Teukolsky equation (or as follows earlier work~\cite{Brill_1972}). Nevertheless, by abstaining from full separation of variables we support subsequent investigation of Lorenz gauge Kerr perturbations via elliptic PDEs. Prior self-force calculations based on hyperbolic PDEs have routinely used the scalar case for development of techniques~\cite{Haas_2007,Barack_2007b,Barack_2007c,Vega_2008,Vega_2009,Canizares_2009,Canizares_2010,Thornburg_2010,Vega_2011,Dolan_2011,Thornburg_2017,Thornburg_2020b}. Our formulation of the scalar field equation in Kerr spacetime in terms of $m$-modes closely follows that of Dolan~et~al.~\cite{Dolan_2011} (except we enter the frequency domain), see Sec.~\ref{sec:PDE}. For self-force regularization we adopt the effective source method~\cite{Barack_2007b,Barack_2007c,Vega_2008,Vega_2009,Vega_2011,Wardell_2012}, see Sec.~\ref{sec:Seff}. Our numerical approach is based on a finite difference representation of the elliptic PDE as a matrix equation, see Sec.~\ref{sec:num}. Our results are consistent with prior Kerr scalar self-force calculations, see Sec~\ref{sec:resu}; in those comparisons our methods are distinguished by their purposeful applicability to Kerr gravitational perturbations in Lorenz gauge. 

\section{Elliptic PDE representation of the Kerr scalar field equation}
\label{sec:PDE}

We consider a scalar field $\Phi$ governed by the massless Klein-Gordon equation in Kerr spacetime
\begin{align}
\label{eq:klein}
& g^{\alpha\beta} \nabla_\alpha \nabla_\beta \, \Phi = -4\pi \rho \, ,
\end{align}
where $\nabla_\alpha$ denotes a covariant derivative, $\rho$ is the scalar charge density, and $g^{\alpha\beta}$ is the inverse Kerr metric with mass parameter $M$ and specific angular momentum parameter $a$. We adopt Boyer-Lindquist coordinates ($t$, $r$, $\theta$, $\phi$). The scalar charge density describes a point charge with spatial position ($r_p$, $\theta_p$, $\phi_p$) and four-velocity $u^\alpha$
\begin{align}
\rho = \frac{q}{u^t\sqrt{-g}}\, \delta(r-r_p)\, \delta(\theta-\theta_p)\, \delta(\phi-\phi_p)\, ,
\end{align}
where $g = -\sin^2{\theta}\,(r^2+a^2\sin^2{\theta})^2$ is the Kerr metric determinant and $q$ is the scalar charge. Generally, $r_p$, $\theta_p$, and $\phi_p$ might all depend on time. For simplicity, we specialize to circular equatorial motion by requiring $r_p = r_0$, $\theta_p = \pi/2$, and $\phi_p = \Omega\,t$. The relationships between $u^t$, $\Omega$, and $r_0$ follow from standard geodesic analysis involving constants of motion~\cite{Carter_1968}
\begin{align}
\Omega &= \frac{1}{a+\sqrt{r_0^3/M}} \, ,
\\
u^t &= \frac{a+\sqrt{r_0^3/M}}{\sqrt{r_0^3/M-3r_0^2+2a\sqrt{r_0^3/M}}} \, .
\end{align}
In this treatment the sign of $a$ determines whether the spin and orbital angular momentum are aligned or antialigned ($a>0$ is prograde, $a<0$ is retrograde). We only consider orbits with $r_0$ greater than or equal to the radius of the innermost stable circular orbit, $r_\text{ISCO}$~\cite{Bardeen_1972}
\begin{align}
& 0 = 1 - \frac{6M}{r_\text{ISCO}} + 8a\sqrt{\frac{M}{r_\text{ISCO}^3}}-\frac{3a^2}{r_\text{ISCO}^2} \, ,
\end{align}
where $r_\text{ISCO} \le 6M$ for prograde orbits and $r_\text{ISCO} \ge 6M$ for retrograde orbits.

Our approach involves separating the $\phi$ and $t$ variables
\begin{align}
\label{eq:modes}
&\Phi(t,r,\theta,\phi) = \frac{1}{r} \sum_{m=-\infty}^{\infty}\Psi_m(r,\theta)\;e^{im\Delta\phi(r)}e^{im(\phi-\Omega t)} \, ,
\end{align}
where $\Psi_m$ represents an $m$-mode of the scalar field and we have assumed circular motion (although, noncircular motion may be similarly accessible, see Sec.~\ref{sec:conc} for discussion). Following Dolan~et~al.~\cite{Dolan_2011}, we have introduced what is effectively a height function $\Delta\phi(r)$ to the azimuthal coordinate to address singular behavior near the event horizon
\begin{align}
\label{eq:dphi}
\Delta\phi(r) = \frac{a}{r_+ - r_-} \ln\left( \frac{r-r_+}{r-r_-} \right) ,
\end{align}
where $r_\pm = M\pm\sqrt{M^2-a^2}$ are the inner ($-$) and outer ($+$) horizons of the Kerr black hole. One benefit of this $\phi$ transformation involves simplification of boundary conditions imposed near $r=r_+$, see Sec.~\ref{sec:BC}. Enforcing orthogonality of the sinusoidal azimuthal harmonics provides the inverse relation
\begin{align}
\label{eq:m-inverse}
\Psi_m(r,\theta) = \frac{r}{2\pi} e^{-im\Delta\phi(r)} \int_0^{2\pi} \Phi(t,r,\theta,\phi)\;e^{-im(\phi-\Omega t)} \, d\phi \, .
\end{align}
Substituting Eq.~\eqref{eq:modes} into Eq.~\eqref{eq:klein} reveals how each $\Psi_m$ is governed by a PDE involving $r$ and $\theta$ derivatives
\begin{align}
\label{eq:box}
&\square_m \equiv -m^2\Omega^2+\frac{4am^2\Omega Mr}{\Sigma^2}-\frac{\left(r^2+a^2\right)^2}{\Sigma^2}\frac{\partial^2}{\partial r_*^2} \notag
\\&\;\;\;\; -\frac{2iamr\left(r^2+a^2\right)-2a^2\Delta}{r\Sigma^2}\frac{\partial}{\partial r_*} -\frac{\Delta}{\Sigma^2}\Bigg(\frac{\partial^2}{\partial\theta^2}
\\&\;\;\;\; +\cot{\theta}\frac{\partial}{\partial\theta}-\frac{m^2}{\sin^2{\theta}}-\frac{2M}{r}\Big(1-\frac{a^2}{Mr}\Big)-\frac{2iam}{r}\Bigg) \, , \notag
\\
\label{eq:ellipticPDE}
& \square_m \, \Psi_m = S_m \, ,
\end{align}
where $\Delta \equiv r^2-2Mr+a^2$, $\Sigma^2 \equiv (r^2+a^2)^2 -a^2 \Delta \sin^2{\theta}$, $S_m$ is an appropriately defined $m$-mode of $\rho$, and $r_*$ is the tortoise coordinate
\begin{align}
&\frac{dr_*}{dr} = \frac{r^2+a^2}{\Delta} ,
\\
&r_* = r+\frac{2M}{r_+ - r_-} \left( r_+ \ln\Big( \frac{r-r_+}{2M}\Big) - r_- \ln\Big( \frac{r-r_-}{2M} \Big) \right) \, .
\end{align}
Although differential operators involving $\,\square\,$ often refer to hyperbolic PDEs, our $\square_m$ is an elliptic PDE operator (notice that the coefficient of $\partial^2/\partial r_*^2$ has the same sign as the coefficient of $\partial^2/\partial\theta^2$).

Before pursuing numerical solutions for $\Psi_m$ and calculating the self-force, we must confront the need for local regularization. Approaching the small body, each $m$-mode diverges logarithmically. While certain dissipative aspects of the self-force can be inferred from asymptotic fluxes, this infinite local behavior obscures the self-interaction that uniquely enables determination of the conservative part of the self-force. We adopt the effective source method to overcome this obstacle.

\section{Regularization via the effective source method}
\label{sec:Seff}

Our regularization method is based on the decomposition of $\Phi$ into singular and regular pieces defined by Detweiler and Whiting~\cite{Detweiler_2003}
\begin{align}
\Phi = \Phi^\text{S} + \Phi^\text{R} \, .
\end{align}
Here we provide a qualitative summary of certain relevant features of $\Phi^\text{S}$ and $\Phi^\text{R}$; see~\cite{Poisson_2004} for a thorough presentation. $\Phi^\text{S}$ is an inhomogeneous solution of the scalar field equation with a vanishing self-force contribution. Therefore, $\Phi^\text{R}$ must be a homogeneous solution of the scalar field equation that is entirely responsible for the self-force $F^\text{self}_\alpha$
\begin{align}
F^\text{self}_\alpha = q \, \nabla_\alpha \Phi^\text{R} \, ,
\end{align}
where the gradient is evaluated at the position of the small body. Naturally, we seek $\Phi^\text{R}$ as a means to calculate $F^\text{self}_\alpha$. The key to finding $\Phi^\text{R}$ is to first find $\Phi^\text{S}$ and $\Phi$. For $\Phi^\text{S}$ to simultaneously satisfy the inhomogeneous field equation and cause no force on $q$, it requires an inherently local definition. This local character precludes any sort of numerical approach that would pursue a global boundary condition to determine $\Phi^\text{S}$ or $\Phi^\text{R}$ directly. Rather, $\Phi^\text{S}$ is accessible through a local expansion~\cite{Detweiler_2003b,Haas_2006,Heffernan_2012,Heffernan_2014}
\begin{align}
\label{eq:sing}
\Phi^\text{S} \sim \frac{A_{1}}{s} + A_{2} + A_{3} \, s  + A_{4} \, s^2 + \mathcal{O}(s^3) \, ,
\end{align}
where $s$ is some appropriate measure of distance from the small body, and each coefficient $A_{j}$ characterizes features such as how $\Phi^\text{S}$ depends on the direction of approach toward the small body. In a coordinate representation (which is convenient in numerical applications), quantifying directional dependence and/or distance involves mathematical ingredients such as $r-r_0$, $\theta-\pi/2$, and $\phi-\Omega t$. Physically, $\Phi^\text{S}$ mimics a tidally distorted Coulomb potential (according to the small body's local rest frame).

In practice, expansions of $\Phi^\text{S}$ are truncated at a certain order; we use $\Phi^\mathcal{P}_{(n)}$ to represent our truncated approximation of the singular field
\begin{align}
\label{eq:truncate}
\Phi^\mathcal{P}_{(n)} = \Phi^\text{S} + \mathcal{O}(s^{n-1}) \, ,
\end{align}
where $n$ is the truncation order. This produces an approximation of the regular field $\Phi^\mathcal{R}$, which is accessible to calculate $F^\text{self}_\alpha$
\begin{align}
\label{eq:regular}
\Phi^\mathcal{R} \equiv \Phi - \Phi^\mathcal{P}_{(n)} \, .
\end{align}
Besides targeting a certain $n$, which does influence practical numerical features, there is considerable ambiguity in deciding how to truncate the expansion that determines $\Phi^\mathcal{P}_{(n)}$. Two distinct versions of $\Phi^\mathcal{P}_{(n)}$ may be equally valid if their difference is negligible compared to the truncation error. This ambiguity can be leveraged in pursuit of desirable properties. Although $\Phi^\text{S}$ is defined locally, $\Phi^\mathcal{P}_{(n)}$ may be numerically accessible globally if it is designed to avoid a vanishing denominator. Similarly, it is convenient to define $\Phi^\mathcal{P}_{(n)}$ so that $\phi-\Omega t$ appears solely within smooth periodic functions. These favorable properties are realized by the implementation of $\Phi^\mathcal{P}_{(n)}$ designed by Wardell and co-workers~\cite{Wardell_2012,Wardell_website,Wardell_github} that is used in this work. Our adopted $\Phi^\mathcal{P}_{(n)}$ has a truncation order of $n=4$, which is sufficient for $\Phi^\mathcal{R}$ to exhibit a smooth gradient at the position of the small body.

Although Eq.~\eqref{eq:regular} defines $\Phi^\mathcal{R}$, it is not practical for numerical calculations because both terms are singular. However, the effective source method provides $\Phi^\mathcal{R}$ as a solution of the scalar field equation with a well-behaved source derived from $\Phi^\mathcal{P}_{(n)}$. Here we illustrate the effective source method from a viewpoint where the $\phi$ and $t$ variables have already been separated. Consider definitions of $\Psi_m^\mathcal{P}$ and $\Psi_m^\mathcal{R}$ that respectively follow from substituting $\Phi^\mathcal{P}_{(4)}$ and $\Phi^\mathcal{R}$ into Eq.~\eqref{eq:m-inverse}
\begin{align}
\Psi_m = \Psi_m^\mathcal{P} + \Psi_m^\mathcal{R} \, .
\end{align}
Knowing that $\Psi_m$ is governed by the elliptic PDE defined in Eq.~\eqref{eq:ellipticPDE}, we seek a field equation governing $\Psi_m^\mathcal{R}$
\begin{align}
& \square_m \, \Psi_m = \, \square_m \left( \Psi_m^\mathcal{P} + \Psi_m^\mathcal{R} \right) = S_m \, .
\end{align}
Because $\Psi_m^\mathcal{P}$ is known from a local expansion, we interpret its contribution as affecting the source governing $\Psi_m^\mathcal{R}$
\begin{align}
\label{eq:Seff}
& \square_m \, \Psi_m^\mathcal{R} = S_m - \square_m \Psi_m^\mathcal{P} \equiv S_m^\text{eff} \, ,
\end{align}
where $S_m^\text{eff}$ is known as the effective source (or perhaps an $m$-mode of the effective source). Although $\Psi_m^\mathcal{P}$ is infinite, $S_m^\text{eff}$ is finite everywhere because $\Psi_m^\mathcal{P}$ approximately satisfies Eq.~\eqref{eq:ellipticPDE} such that the Dirac deltas cancel. Equation~\eqref{eq:Seff} alone is not sufficient to determine the self-force because $\Psi_m^\mathcal{R}$ does not satisfy a predictable boundary condition. However, $\Psi_m$ does satisfy a well-defined boundary condition as the retarded solution. Therefore, we seek $\Psi_m^\mathcal{R}$ only in a localized region surrounding the small body and we adjust our calculation to find $\Psi_m$ elsewhere. Consider a worldtube around the small body with bounds $r_A < r_0 < r_B$ and $\theta_A < \pi/2 < \theta_B$ ($\phi$ bounds would not be helpful in describing each $m$-mode). Depending on position relative to the worldtube, our comprehensive strategy involves
\begin{align}
& \square_m \, \Psi_m = 0 \, , \qquad\qquad\, \text{outside worldtube} \notag
\\
\label{eq:tube_strat}
& \square_m \, \Psi_m^\mathcal{R} = S_m^\text{eff} \, , \qquad\;\;\;\; \text{inside worldtube}
\\
&\Psi_m^\mathcal{R} = \Psi_m  - \Psi_m^\mathcal{P} \, ,
 \qquad\, \text{across worldtube} \; . \notag
\end{align}

Equipped with each $\Psi_m^\mathcal{R}$, the self-force follows from applying Eq.~\eqref{eq:modes} to reconstruct $\Phi^\mathcal{R}$ as a sum over $m$ and calculating the gradient of each $m$-mode evaluated at the particle
\begin{align}
\label{eq:msum}
& F^\text{self}_\alpha = \sum_{m=0}^{\infty} F^m_\alpha \, ,
\\
& F^m_\alpha \equiv \begin{cases} 
      q\,\nabla_\alpha ( \Psi^\mathcal{R}_m/r ) \, , & m = 0 \\
      2 q\, \text{Re}\left[ \nabla_\alpha ( \Psi^\mathcal{R}_m \, e^{im\Delta\phi}e^{im(\phi-\Omega t)}/r ) \right] , & m > 0 \, .
   \end{cases} \notag
\end{align}
In practice, the sum is truncated after a finite number of $m$-modes. The effective source implementation we have adopted accommodates $m\le 20$. With an effective source expansion characterized by $n=4$, truncating the sum over $m$ introduces an error for the conservative part of $F^\text{self}_\alpha$ of size $\sim m_\text{max}^{-4}$; this might seem to imply a negligible error with our $m_\text{max}\simeq 20$, except we find that an especially large coefficient amplifies this truncation error to what would be a limiting factor. Thankfully, usage of $m$-mode regularization parameters~\cite{Heffernan_2014} strengthens our accuracy by accelerating convergence, see Sec.~\ref{sec:resu}.

\section{Numerical techniques}
\label{sec:num}

\subsection{Second-order finite difference method}

We calculate numerical solutions of Eq.~\eqref{eq:tube_strat} through a second-order finite difference method. Our discretized numerical domain is a rectangular grid with fixed $\Delta r_*$ and $\Delta\theta$. The worldtube is centered at $r=r_0$ and $\theta = \pi/2$. Through numerical experimentation we find that a worldtube diameter of $15 M$ (according to $r_*$) and polar width of $\pi/3$ are favorable to minimize steep gradients near the small body. We determine $\Delta r_*$ and $\Delta\theta$ in part by ensuring that an odd number of spaces span the worldtube. Besides the obvious polar domain, $0\le\theta\le\pi$, we introduce asymptotic $r_*$ boundary positions, $r_*^\text{min}\le r_* \le r_*^\text{max}$, sufficiently distant that radiative boundary conditions are approximately valid. Because $\Delta r_*$ was determined according to the worldtube diameter, we require that an integer number of worldtube diameters fit between $r_*^\text{min}$ and $r_*^\text{max}$ so that resolution improvements can be accommodated with fixed domain boundaries. Additionally, we fix the ratio of $r_*$ and $\theta$ grid spacings at $\Delta r_*/\Delta \theta = 45 M/\pi$ (determined through experimentation) to reduce degrees of freedom when conducting resolution tests. As Eq.~\eqref{eq:tube_strat} suggests, we aim to solve for two different types of fields; the retarded field $\Psi_m$ outside the worldtube and the regular field $\Psi_m^\mathcal{R}$ inside the worldtube. We assemble each unknown value at each grid point into a vector of unknowns with indices indicating the associated spatial position: $\vec{\Psi} \equiv (\Psi_{m00},\Psi_{m01},\Psi_{m02}, \,\dots\, , \Psi_{m10},\Psi_{m11}, \,\dots\, , \Psi_{m(i,j)} , \,\dots\,)$. The first index $i$ specifies the radial position: $r_{*(i)} = r_*^\text{min} + i\,\Delta r_*$; the second index $j$ specifies the polar position: $\theta_{(j)} = j \, \Delta\theta\,$.

We approximately represent Eq.~\eqref{eq:tube_strat} by replacing partial derivatives with their equivalent finite difference expression at that position
\begin{align}
\label{eq:diff}
\frac{\partial \Psi_{m(i,j)}}{\partial r_*}  &\rightarrow \frac{\Psi_{m(i+1,j)} -\Psi_{m( i-1,j)}}{2 \Delta r_*} \, ,
\\
\frac{\partial^2 \Psi_{m(i,j)}}{\partial r_*^2} &\rightarrow \frac{\Psi_{m(i+1,j)} -2\Psi_{m(i,j)}+\Psi_{m(i-1,j)}}{\Delta r_*^2} \notag \, ,
\\
\frac{\partial \Psi_{m(i,j)}}{\partial \theta} &\rightarrow \frac{\Psi_{m(i,j+1)} -\Psi_{m(i,j-1)}}{2 \Delta \theta} \notag \, ,
\\
\frac{\partial^2 \Psi_{m(i,j)}}{\partial \theta^2} &\rightarrow \frac{\Psi_{m(i,j+1)} -2\Psi_{m(i,j)}+\Psi_{m(i,j-1)}}{\Delta \theta^2} \, . \notag 
\end{align}
Note that Eq.~\eqref{eq:diff} applies to $\Psi_m$ when all five referenced positions are located outside the worldtube; further analysis involving $\Psi_m^\mathcal{R}$ is necessary when positions within the worldtube are referenced. Each grid point where we seek a field value has an associated equation. For elliptic PDEs (such as ours), all unknowns are universally coupled in a way that compels a fixed global solution; this is in contrast to parabolic or hyperbolic PDEs that would involve initial data and evolution. For each $m$-mode, we express the finite difference equations associated with our elliptic PDE in matrix form to leverage the power of computational linear algebra
\begin{align}
\label{eq:linsys}
\mathbf{M} \, \vec{\Psi} = \vec{S} \, ,
\end{align}
where the matrix $\mathbf{M}$ contains the finite difference coefficients (boundary conditions also influence $\mathbf{M}$, see Sec.~\ref{sec:BC}) and the vector $\vec{S}$ characterizes inhomogeneous features such as $S_\text{eff}^m \,$. 

\begin{figure}
\includegraphics[width=3.3in]{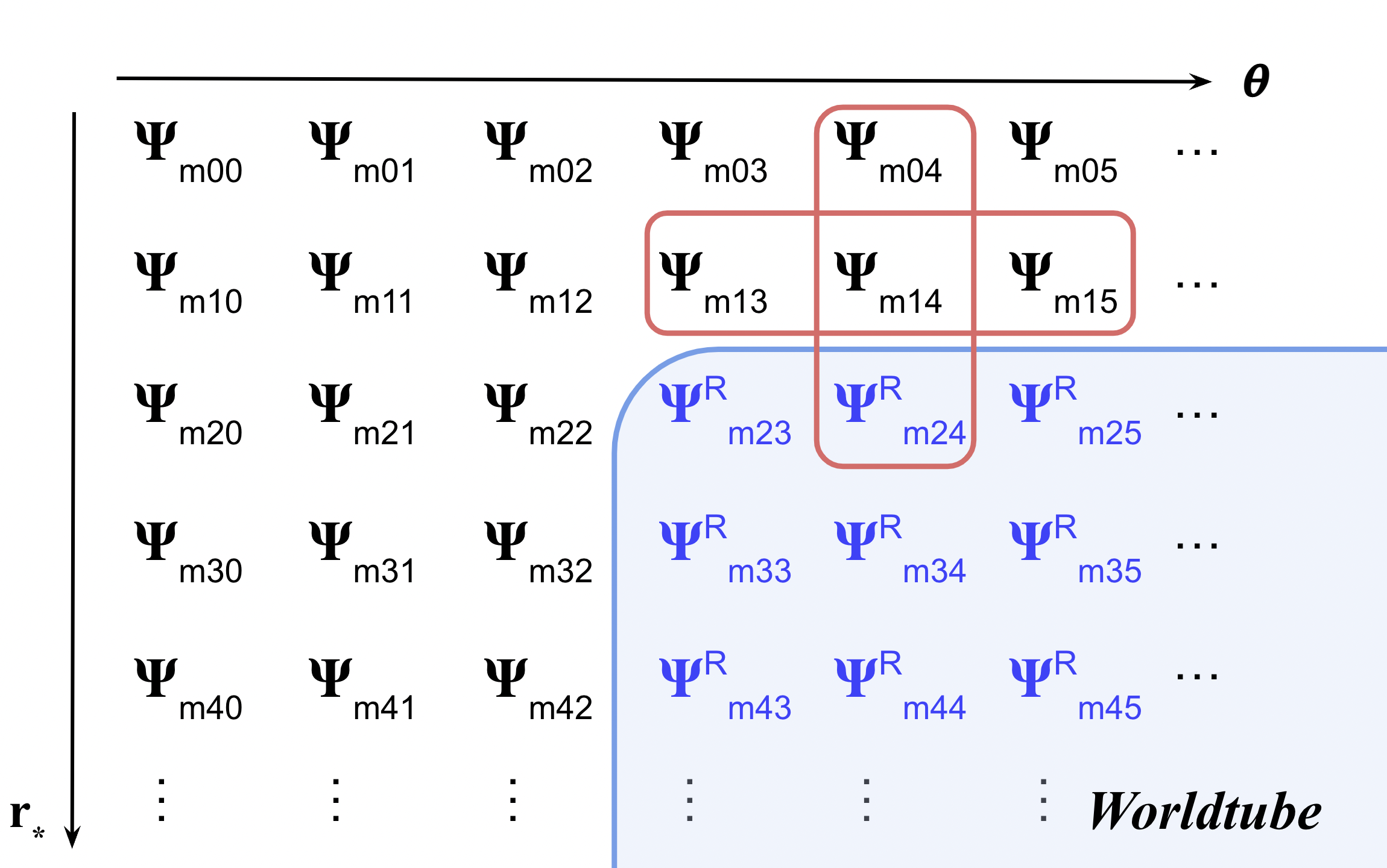}
\caption{\label{fig:cross} Example of a finite difference stencil involving fields inside and outside the worldtube. Notice that $\Psi_{m14}$ has three neighboring fields located outside the worldtube and one inside the worldtube. According to Eq.~\eqref{eq:tube_strat}, we transform between $\Psi_m$ and $\Psi_m^\mathcal{R}$ using $\Psi_m^\mathcal{P}$ as appropriate.}
\end{figure}

Consider a case involving positions within the worldtube. If all five positions referenced in Eq.~\eqref{eq:diff} are within the worldtube, it is sufficient to incorporate the appropriate effective source value and replace $\Psi_m$ with $\Psi_m^\mathcal{R}$ everywhere in Eq.~\eqref{eq:diff}. However, when some referenced positions are outside the worldtube while others are inside, we transform between $\Psi_m$ and $\Psi_m^\mathcal{R}$ using $\Psi_m^\mathcal{P}$ as necessary. As an example, consider the finite difference equation associated with the $(i,j)=(1,4)$ position where the finite difference stencil intersects the worldtube as depicted in Fig.~\ref{fig:cross}. Although the finite difference equation ($\square_m \Psi_{m14}=0$) requires $\Psi_m$ at five positions including $(i,j)=(2,4)$, $\Psi_{m24}$ does not appear in the vector of unknowns; rather, $\Psi^\mathcal{R}_{m24}$ appears in the vector of unknowns at that position inside the worldtube. To express the required $\Psi_{m24}$ in terms of the available $\Psi^\mathcal{R}_{m24}$ we use $\Psi_{m24} = \Psi_{m24}^\mathcal{R}+\Psi_{m24}^\mathcal{P}$ in the finite difference equation
\begin{align}
\label{eq:before}
& A\,  \frac{\Psi_{m24}^\mathcal{R}+\Psi_{m24}^\mathcal{P}-2\Psi_{m14}+\Psi_{m04}}{\Delta r_*^2}
\\
&\qquad\;\;\;\; + B\, \frac{\Psi_{m24}^\mathcal{R}+\Psi_{m24}^\mathcal{P}-\Psi_{m04}}{2\Delta r_*} + \dots = 0 \, , \notag
\end{align}
where $A$ is the coefficient of $\frac{\partial^2\Psi_m}{\partial r_*^2}$ and $B$ is the coefficient of $\frac{\partial\Psi_m}{\partial r_*}$ according to Eq.~\eqref{eq:ellipticPDE}. Because $\Psi_{m24}^\mathcal{P}$ is known we include it in $\vec{S}$ by moving it to the right-hand side,
\begin{align}
\label{eq:after}
& A\, \frac{\Psi_{m24}^\mathcal{R}-2\Psi_{m14}+\Psi_{m04}}{\Delta r_*^2}
\\
&\qquad + B\, \frac{\Psi_{m24}^\mathcal{R} -\Psi_{m04}}{2\Delta r_*}+ \dots = -A\, \frac{\Psi_{m24}^\mathcal{P}}{\Delta r_*^2}-B\, \frac{\Psi_{m24}^\mathcal{P}}{2\Delta r_*} \, . \notag
\end{align}  
Notice that $\mathbf{M}$ is not affected by this transformation. By applying this logic to each case where the finite difference stencil intersects the worldtube, the jumps from $\Psi_{m}$ to $\Psi_{m}^\mathcal{R}$ are accommodated by carefully inserting $\Psi_{m}^\mathcal{P}$ into $\vec{S}$ at the worldtube boundaries.

\subsection{Boundary conditions}
\label{sec:BC}

The $\theta$ boundary conditions follow from two physical requirements: $\Phi$ must be continuous and differentiable away from the particle. At $\theta=0$ and $\theta=\pi$, the continuity requirement implies a condition for $\Psi_{m\ne 0}$ and the differentiability requirement implies a condition for $\Psi_{m=0}$~\cite{Barack_2007b}
\begin{align}
\label{eq:theta}
\Psi_{m \neq 0}\Big|_{\theta=0,\pi} &= \; 0,
\\
\label{eq:thetab}
\frac{\partial}{\partial \theta} \Psi_{m=0} \Big|_{\theta=0,\pi} &= \; 0 .
\end{align}
The justification for why these conditions governing each $\Psi_m$ achieve a continuous and differentiable $\Phi$ is related to how each $m$-mode implies associated $\phi$ dependence (and how that $\phi$ behavior affects $\Phi$ approaching the poles). For the $m=0$ condition, we use second-order one-sided finite difference equations to approximate the $\theta$ derivative
\begin{align}
\label{eq:theta2}
&\frac{-3 \Psi_{m=0}\big|_{\theta=0} + 4 \Psi_{m=0}\big|_{\theta = \Delta \theta} -\Psi_{m=0}\big|_{\theta=2\Delta \theta}}{2\Delta\theta} \simeq 0 \, ,
\\
&\frac{3 \Psi_{m=0}\big|_{\theta=\pi} - 4 \Psi_{m=0}\big|_{\theta = \pi - \Delta \theta} +\Psi_{m=0}\big|_{\theta=\pi - 2\Delta \theta}}{2\Delta\theta} \simeq 0 \, . \notag
\end{align}

The $r_*$ boundary conditions follow from identifying features exhibited by the retarded solution. Specifically, we demand asymptotic wave propagation directed away from the source (either toward decreasing $r_*$ near $r_*\simeq -\infty$ or toward increasing $r_*$ near $r_*\simeq +\infty$). Consider the behavior of $\Psi_m$ near $r=r_+$. It is well known that near-horizon wave propagation in Kerr spacetime involves nontrivial dispersion that is responsible for extraordinary phenomena such as superradiance. By substituting $r=r_+$ into Eq.~\eqref{eq:ellipticPDE} (which causes the $\theta$ derivatives to vanish) we are able to describe how $\Psi_m$ behaves near the horizon
\begin{align}
\label{eq:horWave}
\Psi_m \, e^{-im\Omega t} \Big|_{r_*\simeq -\infty} = f(\theta) \,  e^{i(k^{\pm} r_*-m\Omega t)} \, ,
\end{align}
where $k^{\pm}$ represents the wave numbers associated with waves traveling in the increasing ($+$) or decreasing ($-$) $r_*$ direction
\begin{align}
& k^+ = m\Omega - \frac{am}{r_+} \, , \notag
\\
& k^- = -m\Omega \, .
\end{align}
Naturally, $k^-$ describes the retarded solution. The simplicity of this result derived from Eq.~\eqref{eq:ellipticPDE} is perhaps surprising; this alignment of our desired downgoing solution with the more naive Schwarzschild behavior is a feature of transforming $\phi$ according to $\Delta \phi(r)$ as described in Eqs.~\eqref{eq:modes} and~\eqref{eq:dphi}. Analysis of Eq.~\eqref{eq:horWave} and its $r_*$ derivative provides the near-horizon boundary condition
\begin{align}
\label{eq:horBC}
\left( \frac{\partial \Psi_m}{\partial r_*} - i k^- \Psi_m \right) \bigg|_{r_*=r_*^\text{min}} \simeq 0 \, .
\end{align}
We use the same type of one-sided finite difference for $r_*$ boundary conditions as with Eq.~\eqref{eq:theta2}.

The appropriate large $r_*$ behavior more obviously involves a similar Sommerfeld condition
\begin{align}
\label{eq:somm}
\;\;\;\;\;\;\;\; \left( \frac{\partial \Psi_m}{\partial r_*} - i m \Omega \Psi_m \right) \bigg|_{r_*\simeq \infty} \simeq 0 \, .
\end{align}
Unfortunately, unlike the near-horizon case, very distant values of $r_*^\text{max}$ ($\sim 5000M$) would be necessary for Eq.~\eqref{eq:somm} to accurately represent the retarded solution; such an implementation would not be computationally practical. Rather than applying Eq.~\eqref{eq:somm} at $r_* = r_*^\text{max}$, we seek an improved boundary condition based on asymptotic analysis
\begin{align}
\label{eq:expansion}
\Psi_{m}\Big|_{r_*\simeq \infty} = e^{im\Omega r_*} \left( A(\theta) + \frac{B(\theta)}{r_*} + \frac{C(\theta)}{r_*^2} + \mathcal{O}\Big(\frac{1}{r_*^3}\Big) \right) \, .
\end{align}
Although the series coefficients $A(\theta)$, $B(\theta)$, and $C(\theta)$ are unknown, by carefully forming linear combinations of Eq.~\eqref{eq:expansion} and its $r_*$ derivatives we have discovered more sophisticated conditions involving less error for a given $r_*^\text{max}$. Consider linear combinations of Eq.~\eqref{eq:expansion} and its first $r_*$ derivative following Eq.~\eqref{eq:somm}; although some lower-order terms do cancel, nonvanishing higher-order terms in the series produce a residual
\begin{align}
& \frac{\partial \Psi_{m}}{\partial r_*} -i m\Omega \Psi_{m}  = \mathcal{O}\Big(\frac{1}{r_*^2}\Big) \, .
\end{align}
Each higher derivative of Eq.~\eqref{eq:expansion} included in the linear combination can achieve cancellation of the next nonvanishing lowest-order residual term. The following two linear combinations each achieve curtailment of the residual by one additional order
\begin{align}
\label{eq:BC2}
& \frac{\partial^2 \Psi_{m}}{\partial r_*^2} -2 im\Omega \frac{\partial \Psi_{m}}{\partial r_*} - m^2\Omega^2 \Psi_{m} = \mathcal{O}\Big(\frac{1}{r_*^3}\Big) \, ,
\\
& \frac{\partial^3 \Psi_{m}}{\partial r_*^3} -3 im\Omega \frac{\partial^2 \Psi_{m}}{\partial r_*^2} -3m^2\Omega^2\frac{\partial \Psi_{m}}{\partial r_*} +i m^3\Omega^3 \Psi_{m} \notag
\\&\qquad\qquad\qquad\qquad\qquad\qquad\qquad =\mathcal{O}\Big(\frac{1}{r_*^4}\Big) \, .
\label{eq:BC3}
\end{align} 
Because associated numerical implementations represent each residual as zero, it is apparent that minimizing the size of the residual improves the accuracy of the solution associated with a certain boundary condition. Although we did investigate usage of Eq.~\eqref{eq:BC3} with some success, our ultimate version of the large $r_*$ boundary condition involves substituting $r_*^\text{max}$ into Eq.~\eqref{eq:BC2}; this with Eqs.~\eqref{eq:horBC}, \eqref{eq:theta}, and~\eqref{eq:thetab} assembles a comprehensive boundary strategy.

Equipped with boundary conditions and finite difference equations, we are able to calculate $\mathbf{M}$, which is a sparse diagonally dominated square matrix. The linear system described by Eq.~\eqref{eq:linsys} involving $\mathbf{M}$ and $\vec{S}$ uniquely determines $\vec{\Psi}$ for given a given domain discretization and $m$ value. Our numerical implementation is based on Mathematica's ``SparseArray" data type and ``LinearSolve" function. We predicted that a modern iterative solver, such as using ``Method$\rightarrow$Krylov" with LinearSolve, was likely to be most efficient. However, we found that using ``Method$\rightarrow$Pardiso" with LinearSolve was more efficient than any of the iterative solvers available in Mathematica; the associated ``Pardiso" documentation suggests it is based on a direct solver involving factorization of $\mathbf{M}$, except the factorization is inexact so that the direct solution requires a small amount of iterative refinement.
 
\section{Results and convergence}
\label{sec:resu}

\begin{figure}
\includegraphics[width=3.3in]{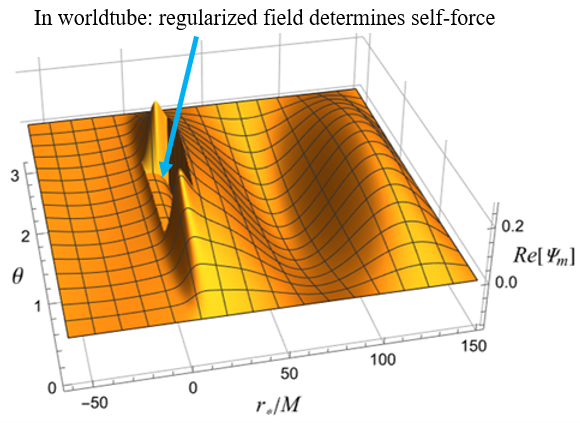}
\caption{\label{fig:sol} Numerical solution of the elliptic PDE governing each $m$-mode of the scalar field. The associated parameters are $a=0.9 M$, $r_0 = 6 M$, and $m=1$. The blue arrow identifies the interior of the worldtube; in that region the numerical solution describes $\Psi_m^\mathcal{R}$, which is responsible for the self-force. Outside the worldtube, the numerical solution describes $\Psi_m$, which is governed by predictable boundary conditions. Across the worldtube we enforce $\Psi_m^\mathcal{R} = \Psi_m - \Psi_m^\mathcal{P}$.}
\end{figure}

\begin{figure}
\includegraphics[width=3.3in]{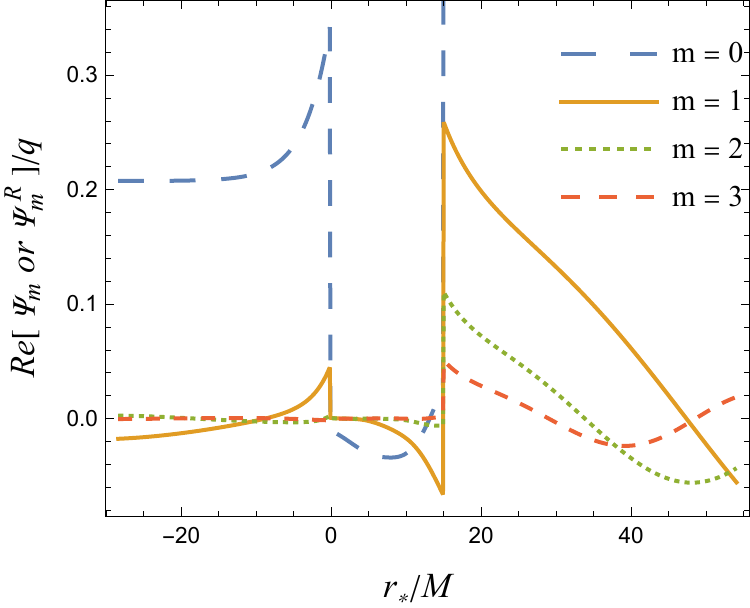}
\caption{\label{fig:radial} Radial cross sections of numerical solutions at $\theta=\pi/2$ for various $m$ values. The associated parameters are $a=0.9 M$ and $r_0 = 6 M$. The worldtube boundaries are identifiable as discontinuities in the numerical solution. We measure the gradient of $\Psi_m^\mathcal{R}$ at the center of the worldtube to assess convergence.}
\end{figure}

Solving the sparse linear system provides $\Psi_m$ and $\Psi_m^\mathcal{R}$ in their respective regions, see Fig.~\ref{fig:sol}. Radial cross sections of the numerical data with $\theta=\pi/2$ for various $m$ values are depicted in Fig.~\ref{fig:radial}. The gradient of $\Psi_m^\mathcal{R}$ at the position of the small body determines each $m$-mode of the self-force. For the dissipative part of the self-force the sum over $m$ converges rapidly. However, for the conservative part of the self-force the sum over $m$ converges more slowly depending on the order of expansion for $\Psi_m^\mathcal{P}$ (or, equivalently, the value of $n$). Generally, the conservative part of the self-force is more challenging to calculate accurately. Therefore, we measure the accuracy of $F^m_r$ specifically to assess any relevant aspect of convergence when calculating $\Psi_m^\mathcal{R}$ numerically. 

\begin{figure}
\includegraphics[width=3.3in]{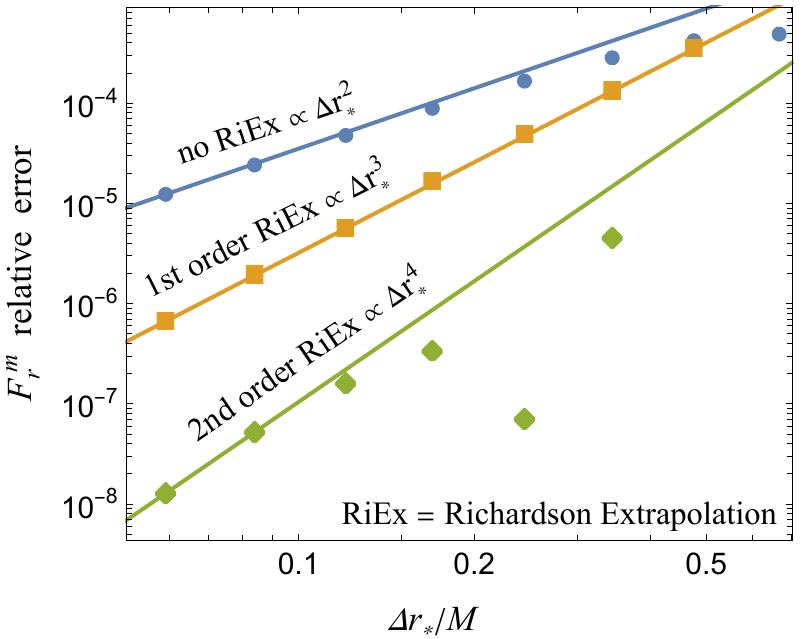}
\caption{\label{fig:res} Measuring convergence with decreasing $\Delta r_*$ and $\Delta\theta$. The associated parameters are $a=0.5 M$, $r_0 = 6 M$, and $m=1$. By default, our second-order finite difference method introduces an error proportional to $\Delta r_*^2$ (or, equivalently, $\Delta \theta^2$), which is rather slow. We accelerate convergence to an acceptable rate with two orders of Richardson extrapolation. The solid lines represent theoretical convergence rates.}
\end{figure}

One obvious aspect of convergence involves ensuring $\Delta r_*$ and $\Delta\theta$ are sufficiently small to minimize discretization error. When pursuing successive resolution refinements, we decrease $\Delta r_*$ and $\Delta\theta$ by the same factor through increasing an integer shared in their denominators. By default, we expect the error of our second-order finite difference method to scale proportionally to $\Delta r_*^2$. However, we use two orders of Richardson extrapolation to accelerate convergence such that our overall performance is equivalent to a fourth-order method, see Fig.~\ref{fig:res}.

\begin{figure}
\includegraphics[width=3.3in]{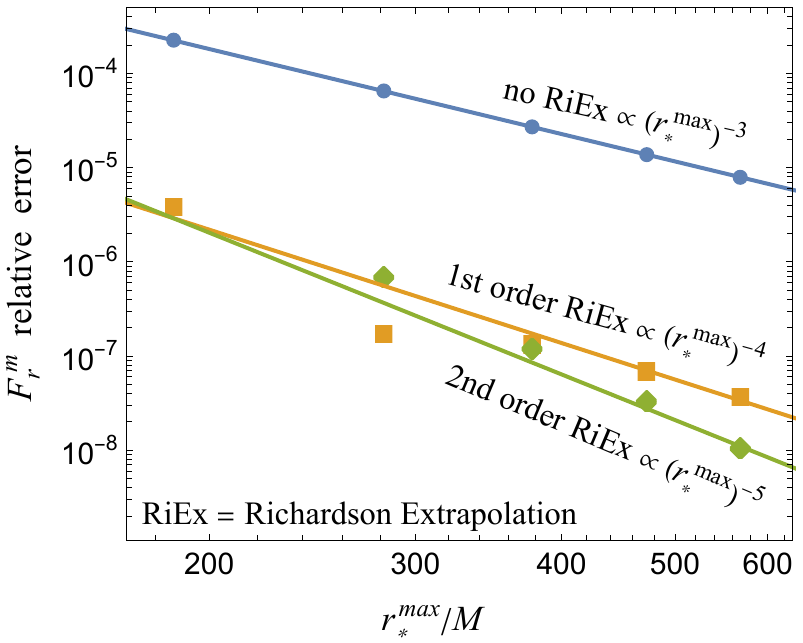}
\caption{\label{fig:domain} Measuring convergence with increasing $r_*^\text{max}$. The associated parameters are $a=0.5 M$, $r_0 = 6 M$, and $m=1$. To ensure our approximate outer boundary condition is not a limiting factor, we successively increase $r_*^\text{max}$ until $F^m_r$ converges. Even with an improved boundary condition that utilizes higher derivatives, we rely on Richardson extrapolation to avoid unreasonably large domain sizes. }
\end{figure}

Another aspect of convergence is related to how our $r_*^\text{min}$ and $r_*^\text{max}$ boundary conditions are only approximate (except at $r_* \simeq \pm \infty$). That approximation improves as $r_*^\text{min}$ is decreased and $r_*^\text{max}$ is increased. We observe rapid convergence with decreasing $r_*^\text{min}$ so that $r_*^\text{min}\simeq -50 M$ is sufficient to avoid limiting errors. In contrast, we evaluate a sequence of increasing $r_*^\text{max}$ values to assess associated convergence. For each $r_*^\text{max}$ we reanalyze discretization errors according to Fig.~\ref{fig:res}. Even with our enhanced boundary condition involving higher $r_*$ derivatives (see Sec.~\ref{sec:BC}), unreasonably high values of $r_*^\text{max}$ are seemingly necessary. Thankfully, we are able to again use Richardson extrapolation to accelerate convergence so that $r_*^\text{max} \simeq 400 M$ is typically sufficient, see Fig.~\ref{fig:domain}.

\begin{figure}
\includegraphics[width=3.3in]{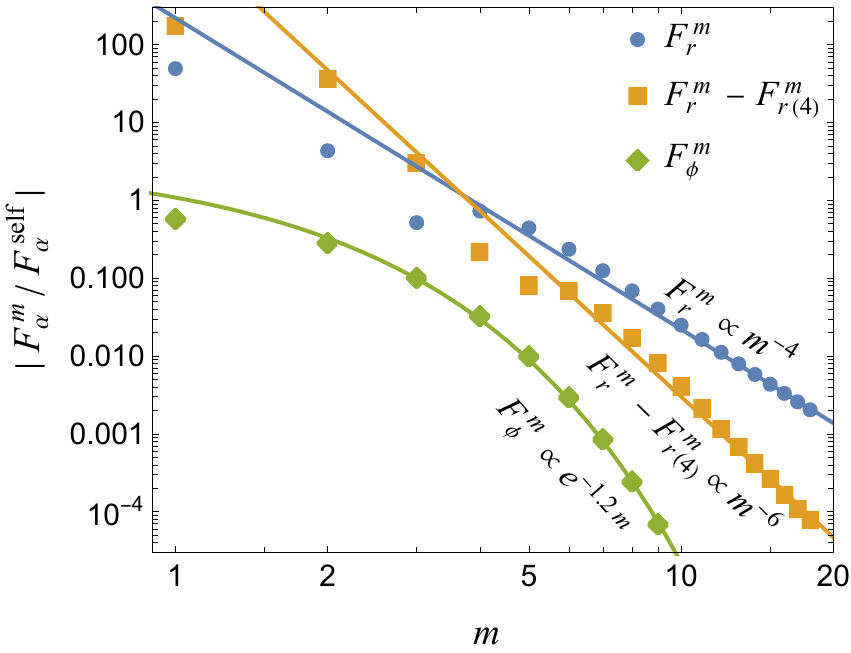}
\caption{\label{fig:msum} Measuring convergence of Eq.~\eqref{eq:msum}. The associated parameters are $a=0.5 M$ and $r_0 = 6 M$. Notice how the dissipative part of the self-force converges exponentially while the conservative part of the self-force converges more slowly. The solid lines represent theoretical convergence rates. We use $m$-mode regularization parameters to accelerate convergence of the conservative part of the self-force.}
\end{figure}

The final aspect of convergence is in applying Eq.~\eqref{eq:msum}. The dissipative part of the self-force converges exponentially as more $m$-modes are included. Unfortunately, the conservative part of the self-force converges at a slower rate, see Fig.~\ref{fig:msum}. Furthermore, some terms of $F^m_r$ are large and positive while others are large and negative which deteriorates our overall accuracy upon cancellation; notice in Fig.~\ref{fig:msum} how the largest term in the sum is bigger than $F^\text{self}_r$ by a factor of $\sim 100$. To improve accuracy, we use $m$-mode regularization parameters~\cite{Heffernan_2014} to essentially include information associated with $m$ values higher than those we calculated numerically. Although the seemingly inaccessible $m$-modes of $q \nabla_r \Phi^\text{R}$ would converge exponentially, our $F^{m}_{r}$ is calculated from the $m$-modes of $q\nabla_r \Phi^\mathcal{R} = q\nabla_r (\Phi-\Phi^\mathcal{P}_{(4)})$; the large-$m$ tail of $F^{m}_{r}$ is an artifact of truncating $\Phi^\mathcal{P}_{(4)}$ and causes slower convergence. Each of those $n=4$ tail terms, $F^m_{r(4)}$, can be represented as the $m$-modes of $q\nabla_r (\Phi^\mathcal{P}_{(5)}-\Phi^\mathcal{P}_{(4)})$. Calculating $F^m_{r(4)}$, which can be evaluated at the particle before $\phi$ integration, is easier than calculating the $m$-modes of $\Phi^\mathcal{P}_{(5)}$ (and its associated higher-order effective source) because the dependence of $\Phi^\mathcal{P}_{(5)}$ on $r$ and $\theta$ would need to be preserved during $\phi$ integration. If $m_\text{max}$ is sufficiently large that $q \nabla_r \Phi^\text{R}$ will have converged, then $F^m_r \simeq F^m_{r(4)}$ when $m>m_\text{max}$. This approximate equivalence is powerful because we have an exact expression for $F^m_{r(4)}$~\cite{Heffernan_2014}, which allows us to include all terms through $m=\infty$ (at $n=4$)
\begin{align}
\label{eq:mreg}
F^\text{self}_r \simeq \sum_{m=0}^{m_\text{max}} F^m_r \;\; + \sum_{m=m_\text{max}+1}^\infty F^{m}_{r(4)} .
\end{align}
Conveniently, the infinite sum in Eq.~\eqref{eq:mreg} has a closed-form expression. $F^{m}_{r(4)}$ is presently the highest-order $m$-mode regularization parameter we are able to access; further enhancements would be achievable if higher-order regularization parameters were also available. It is possible to do numerical fitting for higher-order regularization parameters; we are likely to pursue such numerically determined regularization parameters in future work.

\begin{table}
\caption{Numerical results for $F^\text{self}_r$ compared with prior work. Only significant digits are shown (when present, significant digits beyond the seventh are omitted). Reference~\cite{Warburton_2010} is based on separation of variables via spheroidal harmonics and serves as an accurate benchmark. Reference~\cite{Dolan_2011} is based on hyperbolic PDEs. Notice how, with only one exception, our elliptic PDE method has surpassed or matched the accuracy of Ref.~\cite{Dolan_2011}, which suggests that numerical challenges related to PDEs can be mitigated so long as there are no problematic instabilities.
\label{tbl:Fr}}
\begin{tabular}{| l | l | l | c |}
\hline
\multicolumn{4}{|c|}{Conservative self-force comparisons: $ F_r^{\text{self}}\times(M^2/q^2)$} \\
\hline
  & $\;\;\;\;\;\;\, r_0 = 10M$ & $\;\;\;\;\;\; r_0 = r_{\text{ISCO}}$ & \\
\hline
           & $\msp 4.9400\e5$ & $\msp 9.607\e5$ & This work \\
$a= -0.9M$ & $\msp 4.939995 \e5$ & $\msp 9.607001 \e5$ & Ref. \cite{Warburton_2010} \\
           & $\msp 4.94\e5$ & $\msp 9.607\e5$ & Ref. \cite{Dolan_2011} \\
\hline
           & $\msp 4.1007\e5$ & $\msp 1.1076\e4$ & This work \\
$a=-0.7M$  & $\msp 4.100712 \e5$ & $\msp 1.107625 \e4$ & Ref. \cite{Warburton_2010} \\
           & $\msp 4.10\e5$ & $\msp 1.108\e4$ & Ref. \cite{Dolan_2011} \\
\hline
           & $\msp 3.2894\e5$ & $\msp 1.2751\e4$ & This work \\
$a=-0.5M$  & $\msp 3.28942 \e5$ &  $\msp 1.275170 \e4$ & Ref. \cite{Warburton_2010} \\
	       & $\msp 3.29\e5$ & $\msp 1.275\e4$ & Ref. \cite{Dolan_2011} \\
\hline
           & $\msp 1.3784\e5$ & $\msp 1.677\e4$ & This work \\
$a=\msp 0$ & $\msp 1.378448 \e5$ & $\msp 1.677283 \e4$ & Ref. \cite{Warburton_2010} \\
           & $\msp 1.38\e5$ & $\msp 1.677\e4$ & Ref. \cite{Dolan_2011} \\
\hline
           & $-4.035 \e6$ & $-6.92 \e5$ & This work \\
$a= +0.5M$ & $-4.03517 \e6$ & $-6.922147 \e5$ & Ref. \cite{Warburton_2010} \\
           & $-4.03 \e6$ & $-6.93\e5$ & Ref. \cite{Dolan_2011} \\
\hline
           & $-1.092 \e5$ & $-1.089 \e3$ & This work \\
$a= +0.7M$ & $-1.091819 \e5$ & $-1.088457 \e3$ & Ref. \cite{Warburton_2010} \\
           & $-1.092 \e5$ & $-1.089 \e3$ & Ref. \cite{Dolan_2011} \\
\hline
           & $-1.768 \e5$ & $-1.14 \e2$ & This work \\
$a= +0.9M$ & $-1.768232 \e5$ & $-1.133673 \e2$ & Ref. \cite{Warburton_2010} \\
           & $-1.77 \e5$ & $-1.134 \e2$ & Ref. \cite{Dolan_2011} \\
\hline
\end{tabular}
\end{table}

Table~\ref{tbl:Fr} provides numerical results for $F^\text{self}_r$ and verifies agreement with prior work, Refs.~\cite{Warburton_2010,Dolan_2011}. Table~\ref{tbl:Fphi} provides the same comparison for $F^\text{self}_\phi$ to illustrate that the dissipative self-force is typically not a limiting factor. Reference~\cite{Warburton_2010} is based on separation of variables via spheroidal harmonics and serves as an accurate benchmark. Reference~\cite{Dolan_2011} is based on hyperbolic PDEs. Notice our elliptic PDE method has surpassed or matched the accuracy of Ref.~\cite{Dolan_2011} (with only one exception, see Table~\ref{tbl:Fr}), which suggests that numerical challenges related to PDEs can be mitigated so long as there are no problematic instabilities. Neither of Refs.~\cite{Warburton_2010,Dolan_2011} involve methods that have fully conquered the Lorenz gauge Kerr metric perturbation problem (although, partial success following Refs.~\cite{Dolan_2011,Dolan_2013} has been fruitful~\cite{Isoyama_2014}); we are optimistic that our elliptic PDE method will be able to conquer the gravitational case more thoroughly. Depending on the values of $r_0$ and $a$, our Mathematica implementation that includes automated convergence tests typically consumes $\sim 4$-$128$~GB of peak memory and $\sim 2$-$12$~CPU hours total (including all $m$-modes for a certain $r_0$ and $a$). Because each $m$-mode can be calculated independently, the real world time can be shorter by a factor of $\sim 10$-$20$ (although, our method of parallelization involved assigning each pair of $r_0$ and $a$ to a single core because we considered many orbits).

\begin{table}
\caption{Numerical results for $F^\text{self}_\phi$ compared with prior work. Only significant digits are shown (when present, significant digits beyond the seventh are omitted). Generally, calculating the dissipative part of the self-force is not the limiting factor related to achievable accuracy. Reference~\cite{Dolan_2011} also reported their first insignificant digit (not shown here), which for this table typically aligned with the correct value in a way that was not true for Table~\ref{tbl:Fr} (perhaps their uncertainty estimates for the dissipative self-force were especially careful).
\label{tbl:Fphi}}
\begin{tabular}{| l | l | l | c |}
\hline
\multicolumn{4}{|c|}{Dissipative self-force comparisons: $ F_\phi^{\text{self}}\times(M/q^2)$} \\
\hline
  & $\;\;\;\;\;\;\, r_0 = 10M$ & $\;\;\;\;\;\; r_0 = r_{\text{ISCO}}$ & \\
\hline
           & $- 1.414708 \e3$ & $- 2.188351 \e3$ & This work \\
$a= -0.9M$ & $- 1.414708 \e3$ & $- 2.188351 \e3$ & Ref. \cite{Warburton_2010} \\
           & $- 1.4147 \e3$ & $- 2.1884 \e3$ & Ref. \cite{Dolan_2011} \\
\hline
           & $- 1.356244 \e3$ & $- 2.578045 \e3$ & This work \\
$a=-0.7M$  & $- 1.356244 \e3$ & $- 2.578045 \e3$ & Ref. \cite{Warburton_2010} \\
           & $- 1.3562 \e3$ & $- 2.5780 \e3$ & Ref. \cite{Dolan_2011} \\
\hline
           & $- 1.30227 \e3$ & $- 3.08354 \e3$ & This work \\
$a=-0.5M$  & $- 1.302267 \e3$ &  $- 3.083542 \e3$ & Ref. \cite{Warburton_2010} \\
	       & $- 1.3023 \e3$ & $- 3.0835 \e3$ & Ref. \cite{Dolan_2011} \\
\hline
           & $- 1.185926 \e3$ & $- 5.304232 \e3$ & This work \\
$a=\msp 0$ & $- 1.185926 \e3$ & $- 5.304232 \e3$ & Ref. \cite{Warburton_2010} \\
           & $- 1.1859 \e3$ & $- 5.3042 \e3$ & Ref. \cite{Dolan_2011} \\
\hline
           & $- 1.093493 \e3$ & $-1.1836 \e2$ & This work \\
$a= +0.5M$ & $- 1.093493 \e3$ & $- 1.183567 \e2$ & Ref. \cite{Warburton_2010} \\
           & $- 1.0935 \e3$ & $- 1.1836 \e2$ & Ref. \cite{Dolan_2011} \\
\hline
           & $- 1.062163 \e3$ & $-1.9487 \e2$ & This work \\
$a= +0.7M$ & $- 1.062163 \e3$ & $- 1.948731 \e2$ & Ref. \cite{Warburton_2010} \\
           & $- 1.0622 \e3$ & $- 1.94873 \e2$ & Ref. \cite{Dolan_2011} \\
\hline
           & $- 1.033444 \e3$ & $-4.50 \e2$ & This work \\
$a= +0.9M$ & $- 1.033444 \e3$ & $- 4.508170 \e2$ & Ref. \cite{Warburton_2010} \\
           & $- 1.0334 \e3$ & $- 4.508 \e2$ & Ref. \cite{Dolan_2011} \\
\hline
\end{tabular}
\end{table}

\section{Conclusions and Future Directions}
\label{sec:conc}

We have developed and implemented new methods designed to calculate Lorenz gauge Kerr metric perturbations while avoiding problematic instabilities encountered in prior work~\cite{Dolan_2013}. Our methods involve numerically solving elliptic PDEs with an effective source for each $m$-mode of the perturbation. As proof of concept we applied these methods to calculate the self-force on a scalar charge in a circular orbit around a Kerr black hole. Consistent with our original motivation, we believe the same approach will successfully access the Lorenz gauge gravitational self-force on a compact mass orbiting a Kerr black hole, which has valuable applications related to LISA observations of gravitational waves from EMRIs.

Naturally, achieving an implementation of these methods for Lorenz gauge metric perturbations would be a valuable extension, and we have begun preliminary analysis of that scenario at first order. Anticipating a factor of $\sim 10$ increase in computational cost (through increasing the number of unknowns from one scalar field to ten components of the metric perturbation), it will be beneficial to pursue numerical and/or algorithmic enhancements. Abandoning Mathematica in favor of a more traditional programming language for numerical work is one example of a likely enhancement. Another example is progression of the numerical strategy by pursuing higher-order finite differences or spectral methods. Attaining a domain of fixed size through hyperboloidal slicing and compactification~\cite{Zenginoglu_2008,Zenginoglu_2011,Macedo_2020,Macedo_2022} would be similarly advantageous. Other improvements could include higher-order expansions for the singular field (and effective source) and numerical fitting of higher-order regularization parameters.

Extensions to more comprehensive scenarios include accommodation of noncircular orbital motion and higher order perturbations. Perhaps eccentric and/or inclined motion could be accessible by solving a separate elliptic PDE for each frequency mode, but the nonperfect smoothness of the effective source may impede rapid frequency convergence (the Gibbs phenomenon). In the case of ordinary differential equations, the methods of extended homogeneous solutions~\cite{Barack_2008} and extended particular solutions~\cite{Hopper_2013} have overcome the Gibbs phenomenon to achieve exponential convergence. Discovery of a similar technique to accelerate Fourier convergence with our elliptic PDE approach would be valuable. For higher-order perturbations, first-order Lorenz gauge Kerr self-force calculations would be a promising foundation for second-order Kerr investigations. Although the associated obstacles may be considerable, the methods presented here might become a viable path toward second-order Kerr self-force calculations.

\acknowledgments

T. O. gratefully acknowledges support from the SUNY Geneseo Presidential Fellowship, Proposal Writing Support Award, and Faculty Travel Grant. We thank Barry Wardell for sharing his effective source code, Anna Heffernan, Adrian Ottewill, and Barry Wardell for sharing their $m$-mode regularization parameter expressions, Charles Evans and UNC Chapel Hill for facilitating access to the Longleaf computer cluster, Jonathan Thornburg for communicating about the status of his hyperbolic PDE work, and the anonymous referees for their valuable feedback.

\bibliography{main}

\end{document}